\documentclass[11pt]{article}
\usepackage[margin=2cm]{geometry}
\usepackage{color} 

\usepackage{amsmath,amssymb}

\newcommand{\ba}{\begin{eqnarray}}
\newcommand{\ea}{\end{eqnarray}}
\newcommand{\be}{\begin{equation}}
\newcommand{\ee}{\end{equation}}

\begin{document}

\title{\textbf{Conditions for global minimum in the $A_4$ symmetric 3HDM}}

\author{
Sergio Carrolo \footnote{sergio.carrolo@tecnico.ulisboa.pt},
Jorge C. Rom\~{a}o \footnote{jorge.romao@tecnico.ulisboa.pt},
Jo\~{a}o P. Silva \footnote{jpsilva@cftp.ist.utl.pt},
\\
CFTP, Departamento de Física,\\
Instituto Superior T\'{e}cnico, Universidade de Lisboa, \\
Avenida Rovisco Pais 1, 1049 Lisboa, Portugal \\[5pt]
}

\maketitle

\begin{abstract}
There has been a great interest in three Higgs doublet models (3HDM)
symmetric under an exact $A_4$ symmetry.
We provide the complete analytic necessary and sufficient
conditions for a point to be the global minimum along the neutral directions
in such models,
discussing many of the subtleties involved.
We also present a number of numerical examples, to highlight
those issues.
We then turn to the directions which break electric charge,
presenting a safe analytical sufficient condition
for bounded from below (BFB) potentials.
Based on extensive numerical simulations,
we discuss one conjecture on BFB along charge breaking directions.
\end{abstract}

\section{Introduction}

Since the discovery of a scalar particle with 125GeV at the
LHC \cite{ATLAS:2012yve,CMS:2012qbp},
there has been a strong theoretical effort and experimental search for
extra scalar particles.
A particularly interesting framework is provided by
N Higgs doublet models (NHDM).
For example,
models with three Higgs doublets (3HDM) supplemented
by the $A_4$ discrete symmetry have long been explored in connection
with flavour physics; see, for example,
\cite{Ma:2001dn,Machado:2010uc,deAdelhartToorop:2010nki,Boucenna:2011tj,
GonzalezFelipe:2013xok,GonzalezFelipe:2013yhh,Pramanick:2017wry,Chakrabarty:2018yoy}.

But,
it was only in 2012 that
Degee, Ivanov and Keus carefully addressed the identification of
viable candidates for minima in the $A_4$ 3HDM
\cite{Degee:2012sk}.\footnote{Further developments appeared in
\cite{deMedeirosVarzielas:2017glw}.}
They provide conditions for only two of the solutions of the stationarity
equations to be minima.
Ref.~\cite{deMedeirosVarzielas:2022kbj} revisited this issue,
proposing conditions for two other solutions to be minima.
We have found that some conditions are incorrect, while most
only guarantee that the minimum is local,
not global.
And, in some cases, such local minima occur for potentials
which are in fact unbounded from below.
Indeed, a different but related issue concerns the conditions
for the potential to be bounded from below (BFB).
This turns out to have an interesting impact in our analysis.

As pointed out in Ref.~\cite{Faro:2019vcd},
one must be careful to distinguish conditions for the potential
to be bounded from below along the directions of neutral
vacua (BFB-n), from the conditions that guarantee
that the potential is bounded from below along the
directions of charge breaking vacua (BFB-c).
It turns out that the necessary and sufficient BFB-n conditions
for the $A_4$ have been conjectured in \cite{Ivanov:2020jra}
and proven in \cite{Buskin:2021eig}.
These hinged on the fact that Ref.~\cite{Degee:2012sk}
fully mapped the orbit space of the neutral directions
of the $A_4$ 3HDM.
At the moment,
there is no such analysis for the charge breaking (CB) directions.

Here,
we address all these issues.
After summarizing the known results,
we point out in Sec.~\ref{subsec:CC'} that, contrary to previous assumptions, one must
look at five, not four, candidate minima.
In Sec.~\ref{sec:BFB-n},
we re-express the BFB-n conditions in a way that connects directly
with the existence of local minima.
Section~\ref{sec:global-n} has the first main results.
We write the correct conditions for all five points
(candidates for global minima of the $A_4$ 3HDM)
to be local minima.
We then show that, in general, this does \textit{not}
guarantee that such points are global minima,
providing both the relevant conditions and explicit numerical counter-examples.
We end with a summary of the necessary and sufficient conditions for
a point to be the global minimum consistent with BFB-n.
This \textit{assumes} both that the potential will be BFB-c and that
there is no lower-lying CB global minimum.

However,
one must still address the issue of the charge breaking directions:
Does the potential satisfy BFB-c conditions?;
Are there lower-lying CB global minima?
This is what we turn to in Section~\ref{sec:BFB-c}.
We start by developing a safe sufficient condition for the potential
to be BFB-c. We show that, without this condition,
two dangerous situations may occur.
There may be some points that pass BFB-n and are global minima of neutral directions,
but which, nevertheless, are not the global minima of the potential,
since there are lower-lying CB solutions.
There may even be points that
pass the conditions just mentioned
but correspond to potentials unbounded from below along charged directions.
Using extensive numerical simulations, we found that,
after the conditions for BFB-n and of global minima are imposed,
there is only one type of point (identified as point $D$ below)
where we found lower-lying (that is, \textit{true}) CB global minima or
potentials unbounded from below.
We conjecture that this is a general feature.
We present our conclusions in Sec.~\ref{sec:concl}.

\section{The $A_4$-symmetric 3HDM potential}

\subsection{Potential and candidates for global minimum}

The potential for the $A_4$ symmetric three Higgs doublet model (3HDM)
is, in the notation of \cite{Degee:2012sk},
\ba
V_{A_4} &=& 
- \frac{M_0}{\sqrt{3}} \left(\phi_1 ^ \dagger \phi_1
+ \phi_2 ^ \dagger \phi_2 + \phi_3 ^ \dagger \phi_3\right)
+ \frac{\Lambda_0}{3}  \left(\phi_1 ^ \dagger \phi_1
+ \phi_2 ^ \dagger \phi_2 + \phi_3 ^ \dagger \phi_3\right) ^ 2
\nonumber\\*[1mm]
&&
+ \Lambda_1 \left[ 
\left( \textrm{Re}\left\{\phi_1^\dagger\phi_2\right\} \right)^2
+ \left( \textrm{Re}\left\{\phi_2^\dagger\phi_3\right\} \right)^2
+ \left( \textrm{Re}\left\{\phi_3^\dagger\phi_1\right\} \right)^2
\right]
\nonumber\\*[1mm]
&& 
+ \Lambda_2 \left[ 
\left( \textrm{Im}\left\{\phi_1^\dagger\phi_2\right\} \right)^2
+ \left( \textrm{Im}\left\{\phi_2^\dagger\phi_3\right\} \right)^2
+ \left( \textrm{Im}\left\{\phi_3^\dagger\phi_1\right\} \right)^2
\right]
\nonumber\\*[1mm]
&&
+ \frac{\Lambda_3}{3} \left[ (\phi_1 ^ \dagger \phi_1) ^ 2
+ (\phi_2 ^ \dagger \phi_2) ^ 2 + (\phi_3 ^ \dagger \phi_3) ^ 2 
- (\phi_1^\dagger \phi_1) (\phi_2^\dagger \phi_2)
- (\phi_2^\dagger \phi_2) (\phi_3^\dagger \phi_3)
- (\phi_3^\dagger \phi_3) (\phi_1^\dagger \phi_1)\right]
\nonumber\\*[1mm]
&&
+ \Lambda_4 \left[
\textrm{Re}\left\{\phi_1^\dagger\phi_2\right\}\,
	\textrm{Im}\left\{\phi_1^\dagger\phi_2\right\}
+  \textrm{Re}\left\{\phi_2^\dagger\phi_3\right\}\,
	\textrm{Im}\left\{\phi_2^\dagger\phi_3\right\}
+ \textrm{Re}\left\{\phi_3^\dagger\phi_1\right\}\,
	\textrm{Im}\left\{\phi_3^\dagger\phi_1\right\}
\right] \,.
\label{V_A4}
\ea
Setting $\Lambda_4 = 0$, it reduces to the $S_4$ invariant 3HDM potential.
Therefore, the $A_4$ and $S_4$ models have 6 and 5 free parameters, respectively.

Contrary to what seemed implicit in \cite{Degee:2012sk},
and became explicit in subsequent papers using their result
- including, for example, \cite{Ivanov:2014doa,deMedeirosVarzielas:2022kbj} -
there are 5 (and not 4) different vev directions that can lead to a global
minimum. They are:
\begin{align}
\label{eq::symmvevs}
& A:\  \tfrac{v}{\sqrt{2}}(1,0,0)\, , \qquad \qquad B:\ \tfrac{v}{\sqrt{6}}(1,1,1)\, ,
\qquad \qquad D:\ \tfrac{v}{2}(1,e^{i\alpha},0)\, ,\\*[2mm]
& C:\ \tfrac{v}{\sqrt{6}}( \pm 1,\eta,\eta^{-1})\, ,
\qquad \qquad C':\ \tfrac{v}{\sqrt{6}}(\pm 1,\omega,\omega^2)\, ,
\end{align}
where $\eta = e^{i \pi/3}$, $\omega = \eta^2 = e^{2i\pi/3}$, and
$\alpha$ is fixed by the relations
\begin{equation}
  \sin 2\alpha= - \frac{\Lambda_4}{\sqrt{\left( \Lambda_1 - \Lambda_2 \right)^2 + \Lambda_4^2}}\, ,
\quad\quad  \quad \cos 2\alpha
= - \frac{\Lambda_1 - \Lambda_2}{\sqrt{\left( \Lambda_1 - \Lambda_2 \right)^2 + \Lambda_4^2}} \, .
  \label{alpha}
\end{equation}
The minima have been normalized such that $v_1^2 + v_2^2 + v_3^2 =v^2/2$.
Notice that $A_4$ contains $\mathbb{Z}_2 \times \mathbb{Z}_2$,
and that one has invariance under global hypercharge.
As a result,
the potential is invariant under independent sign changes of $\phi_1$,
$\phi_2$, and $\phi_3$, as can be checked from direct
inspection of \eqref{V_A4}.
Thus, the $\pm$ in $C$ and $C'$ is immaterial.

Part of the confusion in the literature as to the number of minima appeared
because $C'$ is not defined in \cite{Degee:2012sk},
although it is clear from their figure 3 that there are two distinct minima;
$C$ and $C'$.
Moreover,
\cite{Ivanov:2014doa} renames $C'$ as $C$ and does not mention
a second possibility.

Why is this relevant? It is relevant because, for a given set of $\Lambda$'s,
$C$ and $C'$ do \textit{not} yield the same value for the potential at that
minimum. That is $V(C) \neq V(C')$; one is the true global minimum and the
other is not.
Adding to the confusion,
starting from $C$:
\be
(-1, \eta, \eta^{-1}) = \eta^{-1} (\omega^2,\omega,1)\, ,
\ee
Since an overall phase is irrelevant,
it would thus seem that one could get to $C'$ simply by transposing
$\phi_1 \leftrightarrow \phi_3$.
However,
transpositions such as this are \textit{not} in $A_4$.
Indeed,
performing $\phi_1 \leftrightarrow \phi_3$ in eq.~\eqref{V_A4},
the potential does not remain invariant.
Fortunately,
the transformed potential is obtained from the initial one
by simply changing $\Lambda_4 \rightarrow -\Lambda_4$.
Thus, the formulae obtained in \cite{Degee:2012sk} for $C$ can be adapted to $C'$
by simply changing the sign of $\Lambda_4$.
Still, we emphasize that, for chosen
$\Lambda$'s, $V(C) \neq V(C')$.

Part of the confusion may be due to the fact that transpositions
such as $\phi_1 \leftrightarrow \phi_3$ \textit{are} a part of
$S_4$.
In that case,
the potential has $\Lambda_4=0$
and the distinction between $C$ and $C'$ disappears.

\subsection{Vevs, mass spectra, and values of the potential}

The spectra for $A_4$ has been studied in \cite{Degee:2012sk},
and here we summarize those results.
We denote the charged scalar pairs by
$H_{i}^\pm$ ($i=1,2$),
and the new neutral scalars (besides the lowest lying SM-like Higgs)
by $H_k$ ($k=1,2,3,4$).
In cases where a mass ordering of the latter is clear,
we denote by $h_i$ ($H_i$) the masses of the lightest (heaviest) neutral scalars.

The SM-like Higgs boson, $H_{SM}$,
always has a mass given by
$m_{H_{SM}}^2 = \frac{2}{\sqrt{3}}M_0$.
Thus, $M_0>0$ for all cases.
This is related to the following fact.
One can use a unitary $3 \times 3$ transformation in the
space of doublets (the so-called Higgs-family space),
to go into some other basis.
A particularly useful basis~\footnote{To be precise, there exist an infinite collection
of Higgs bases, since, once the vev is in the first doublet,
a $(N-1)\times(N-1)$ unitary transformation on the remaining doublets
does not affect the vev.}
is the Higgs basis \cite{Botella:1994cs},
where only the first doublet carries a vev
\cite{Georgi:1978ri,Donoghue:1978cj,Lavoura:1994fv}.
In this basis,
the imaginary part of the neutral component of the
first doublet corresponds always to the neutral 
would-be Goldstone boson $G^0$.
The corresponding
real part of the neutral component of the
first doublet corresponds, in general,
to a neutral Higgs field that will mix with other four
neutral components into the physical (mass eigenstate) neutral scalars.
However,
in the particular case of the 3HDM with an exact $A_4$ symmetry,
for each of the five points of the $A_4$ 3HDM parameter space
identified in \cite{Degee:2012sk} as candidates for global minima,
there is always a transformation into a Higgs basis
where that neutral scalar decouples completely from the remaining four fields.
Of course, this would not be true for minima of a general 3HDM
(in which case, even in the Higgs basis, the five neutral fields other than $G^0$
mix through a $5 \times 5$ mass matrix);
but it is true for the case discussed in \cite{Degee:2012sk}.

\subsubsection*{$\mathbf{A:\ (1, 0, 0)}$}

For the  $(1, 0, 0)$ solution of the stationarity condition,
the magnitude of the vev is given by
\begin{equation}
    v^2 = \frac{\sqrt{3}M_0}{\Lambda_0 + \Lambda_3}.
\label{vev:A}
\end{equation}
We will sometimes use $\Lambda_A=\Lambda_3$.
The spectra is then expressed as
\begin{align}
m_{H^\pm_i}^2 & =
- \frac{v^2}{2} \Lambda_3 \qquad \textrm{(doubly degenerate)}\, ,
\label{mc:A}
\\
m_{h_i}^2 & = \frac{v^2}{4} \left(\Lambda_1 + \Lambda_2 -
2 \Lambda_3 - \sqrt{(\Lambda_1 - \Lambda_2)^2 + \Lambda_4^2} \right)
\qquad \textrm{(doubly degenerate)}\, ,
\label{mn1:A}
\\
m_{H_i}^2 & = \frac{v^2}{4} \left(\Lambda_1 + \Lambda_2
- 2 \Lambda_3 + \sqrt{(\Lambda_1 - \Lambda_2)^2
+ \Lambda_4^2} \right)
\qquad \textrm{(doubly degenerate)}\, .
\label{mn2:A}
\end{align}
The value of the potential at the minimum is
\be
V(A) = -\frac{M_0^2}{4(\Lambda_0 + \Lambda_3)}\, .
\label{V(A)}
\ee

\subsubsection*{$\mathbf{B:\ (1, 1, 1)}$}

For the $(1, 1, 1)$ solution of the stationarity condition,
the magnitude of the vev is given by
\begin{equation}
    v^2 = \frac{\sqrt{3}M_0}{\Lambda_0 + \Lambda_1}.
\label{vev:B}
\end{equation}
We will sometimes use $\Lambda_B = \Lambda_1$.
The charged Higgs spectra has less degeneracies than at point $A$,
and we find
\begin{align}
m_{H^\pm_i}^2 & = \frac{v^2}{2} \left(- \Lambda_1
\pm \frac{\Lambda_4}{2 \sqrt{3}}\right)\, ,
\label{mc:B}
\\
m_{h_i}^2 & = \frac{v^2}{12} \left( - 5 \Lambda_1
+ 3 \Lambda_2 + 2 \Lambda_3 - \sqrt{(- \Lambda_1
+ 3 \Lambda_2 - 2 \Lambda_3)^2 + 12 \Lambda_4^2} \right)
\qquad \textrm{(doubly degenerate)}\, , 
\label{mn1:B}
\\
    m_{H_i}^2 & = \frac{v^2}{12} \left( - 5 \Lambda_1
+ 3 \Lambda_2 + 2 \Lambda_3 + \sqrt{(- \Lambda_1
+ 3 \Lambda_2 - 2 \Lambda_3)^2 + 12 \Lambda_4^2} \right)
\qquad \textrm{(doubly degenerate)}\, .
\label{mn2:B}
\end{align}
The value of the potential at the minimum is
\be
V(B) = -\frac{M_0^2}{4(\Lambda_0 + \Lambda_1)}\, .
\label{V(B)}
\ee

\subsubsection*{$\mathbf{C:\ (1, \eta, \eta^{-1})}$}

For the $(1, \eta, \eta^{-1})$ solution of the stationarity condition,
the magnitude of the vev is given by
\begin{equation}
    v^2 = \frac{\sqrt{3} M_0}{\Lambda_0 + \Lambda_C},
\label{vev:C}
\end{equation}
where
\be
\Lambda_C=
\frac{1}{4} \left( \Lambda_1 + 3 \Lambda_2 + \sqrt{3} \Lambda_4 \right)\, .
\label{Lambda_C}
\ee

The masses are given by
\begin{align}
m_{H_i^\pm} & = -\frac{v^2}{2} 
\left(\Lambda_2 +\frac{1}{2\sqrt{3}} \Lambda_4\right)\, ,
\label{mc:C}
\qquad
-\frac{v^2}{4}  \left(\Lambda_1 + \Lambda_2 +\frac{2}{\sqrt{3}}
\Lambda_4\right) \, .
\\
m_{h_i}^2 & =  \frac{v^2}{12}(A - B)\, ,
\ \ \ \ \ \  m_{H_i}^2  =  \frac{v^2}{12}(A + B)
\ \ \ \textrm{(doubly degenerate)}\, ,
\label{mn:C}
\end{align}
where
\ba
A &=& -8\Lambda_C + 3(\Lambda_1 + \Lambda_2) + 2 \Lambda_3
\nonumber\\
&=&
\Lambda_1 - 3 \Lambda_2 + 2 \Lambda_3 - 2 \sqrt{3} \Lambda_4
\nonumber\\
&=&
(a+c)+(b+c)\, ,
\label{A}
\\*[2mm]
B &=& 
\sqrt{(4\Lambda_C - 3(\Lambda_1 + \Lambda_2) + 2 \Lambda_3)^2
+ 4 (2\Lambda_C + \Lambda_1 -3 \Lambda_2)^2}
\nonumber\\*[1mm]
&=&
\sqrt{(-2 \Lambda_1 + 2 \Lambda_3 + \sqrt{3} \Lambda_4)^2
+
(3 \Lambda_1 - 3 \Lambda_2 +  \sqrt{3} \Lambda_4)^2}
\nonumber\\
&=&\sqrt{(a-c)^2+(b-c)^2}\, ,
\label{B}
\ea
and\footnote{This is a convenient notation introduced in
\cite{deMedeirosVarzielas:2022kbj} for point $C'$;
adapted here to point $C$ with
$\Lambda_4 \rightarrow -\Lambda_4$.}
\be
a=3(\Lambda_1 - \Lambda_2)\, ,
\ \ \ 
b= 2(\Lambda_3 - \Lambda_1)\, ,
\ \ \ 
c=- \sqrt{3} \Lambda_4\, .
\ee
The value of the potential at the minimum is
\be
V(C) = 
-\frac{M_0^2}{4(\Lambda_0 + \Lambda_C)}\, .
\label{V(C)}
\ee

\subsubsection*{$\mathbf{C':\ (1, \omega, \omega^2)}$}

This case can be obtained from the previous one through
$\Lambda_4 \rightarrow - \Lambda_4$.
The vev becomes,
\begin{equation}
    v^2 = \frac{\sqrt{3} M_0}{\Lambda_0 + \Lambda_{C'}},
\label{vev:C'}
\end{equation}
where
\be
\Lambda_{C'}=
\frac{1}{4} \left( \Lambda_1 + 3 \Lambda_2 - \sqrt{3} \Lambda_4 \right)\, .
\label{Lambda'_C}
\ee
The charged Higgs masses are given by
\begin{align}
m_{H_i^\pm} & = -\frac{v^2}{2} 
\left(\Lambda_2 - \frac{1}{2\sqrt{3}} \Lambda_4\right)\, ,
\qquad
-\frac{v^2}{4}  \left(\Lambda_1 + \Lambda_2 - \frac{2}{\sqrt{3}}
\Lambda_4\right) \, .
\label{mc:C'}
\\
m_{h_i}^2 & =  \frac{v^2}{12}(A' - B')\, ,
\ \ \ \ \ \  m_{H_i}^2  =  \frac{v^2}{12}(A' + B')
\ \ \ \textrm{(doubly degenerate)}\, ,
\label{mn:C'}
\end{align}
where
\ba
A' &=& -8  \Lambda_{C'} + 3(\Lambda_1 + \Lambda_2) + 2 \Lambda_3\, ,
\label{A'}
\\
B' &=& 
\sqrt{(4 \Lambda_{C'} - 3(\Lambda_1 + \Lambda_2) + 2 \Lambda_3)^2
+ 4 (2 \Lambda_{C'} + \Lambda_1 -3 \Lambda_2)^2}\, .
\label{B'}
\ea
The value of the potential at the minimum is
\be
V(C') = 
-\frac{M_0^2}{4(\Lambda_0 + \Lambda_{C'})}\, .
\label{V(C')}
\ee

\subsubsection*{$\mathbf{D:\ (1, e^{i\alpha}, 0)}$}

For the $(1, e^{i\alpha}, 0)$ solution of the stationarity condition,
the magnitude of the vev is given by
\begin{equation}
    v^2 = \frac{4 \sqrt{3} M_0}{4 \Lambda_0 + \Lambda_3 - 3 \Tilde{\Lambda}}
    = \frac{\sqrt{3} M_0}{\Lambda_0 + \Lambda_D},
\label{vev:D}
\end{equation}
where we have defined
\begin{equation}
    \Tilde{\Lambda} = \frac{1}{2}
\left( \sqrt{(\Lambda_1 - \Lambda_2)^2 + \Lambda_4^2}
- (\Lambda_1 + \Lambda_2)\right)\, ,
\label{Ltilde}
\end{equation}
and
\be
\Lambda_D = \frac{1}{4} (\Lambda_3 - 3 \Tilde{\Lambda})
=
\frac{3}{8}
\left[
\Lambda_1 + \Lambda_2 + \frac{2}{3} \Lambda_3
-\sqrt{(\Lambda_1 - \Lambda_2)^2 + \Lambda_4^2}
\right]\, .
\label{Lambda_D}
\ee

The corresponding spectra is
\begin{align}
m_{H^\pm_i}^2 & = \frac{v^2 }{2} \Tilde{\Lambda},
\qquad \frac{v^2}{4} (\Tilde{\Lambda} - \Lambda_3)\, ,
\label{mc:D}
\\
m_{H_i}^2 = & 
\frac{v^2}{2} ( \Tilde{\Lambda} + \Lambda_3)\, ,
\qquad 
\frac{v^2}{2}(2 \Tilde{\Lambda} + \Lambda_1 + \Lambda_2)\, ,
\label{mn1:D}
\\*[2mm]
& 
\frac{v^2}{4}
\left[ 
(\Tilde{\Lambda} + \Lambda_1 + \Lambda_2 - \Lambda_3)
\pm
(2 \Tilde{\Lambda} + 3 \Lambda_1 - \Lambda_2)\, 
\sqrt{\frac{\Tilde{\Lambda} + \Lambda_2}{
2 \Tilde{\Lambda} + \Lambda_1 + \Lambda_2}}
\right]\, .
\label{mn2:D}
%
\end{align}
%
Notice that the definition of $\Tilde{\Lambda}$ in Eq.~\eqref{Ltilde}
guarantees that $\Tilde{\Lambda} + \Lambda_1 >0$,
$\Tilde{\Lambda} + \Lambda_2 >0$,
and that
$2 \Tilde{\Lambda} + \Lambda_1 + \Lambda_2 \geq |\Lambda_1-\Lambda_2| >0$.
This implies that the second Eq.~\eqref{mn1:D} is always positive.
In this case there are no degeneracies in the expressions
for the masses of the neutral scalars,
and we have separated by commas the four $m_{H_i}^2$.
The value of the potential at the minimum is
\be
V(D) = 
-\frac{M_0^2}{4\Lambda_0 + \Lambda_3 - 3 \Tilde{\Lambda}}\, .
%
\label{V(D)}
\ee
Alternatively, we may write
\be
V(D) = 
-\frac{M_0^2}{4(\Lambda_0 + \Lambda_D)}\, .
\label{V(D)_2}
\ee

\subsection{\label{subsec:CC'}The necessary distinction between $C$ and $C'$}

As mentioned in the introduction,
the five minima $A$, $B$, $D$, $C$, and $C'$ appear in
figure 3 of \cite{Degee:2012sk}.
But the corresponding $C'$ masses and conditions are
not presented and the fact that the two minima are
\textit{distinct} is absent from \cite{Ivanov:2014doa} ,
and subsequent literature, including 
\cite{deMedeirosVarzielas:2022kbj}.
Imagine that one is looking at a particular point in parameter
space, corresponding to the values
\ba
&&
M_0 = 13.887500 v^2\, ,
\ \ \ 
\Lambda_0 = 2.928220\, ,
\ \ \ 
\Lambda_1 = -0.305014\, ,
\ \ \ 
\Lambda_2 = -1.018350\, ,
\nonumber\\
&&
\Lambda_3 = 1.365140\, ,
\ \ \ 
\Lambda_4 = -1.053270\, .
\ea
After some preliminary numerical calculations,
one might suspect that $C': (1, \omega, \omega^2)$
(unfortunately renamed $C$ in
\cite{Ivanov:2014doa,deMedeirosVarzielas:2022kbj})
is the lowest minimum,
with the value of the potential at the minimum given by
Eq.~\eqref{V(C')} as
\be
V(C') = -0.189506 \times 10^2  v^4\, .
\ee
But, it is \textit{not}.
The correct minimum is $C: (1,\eta, \eta^{-1})$ and the value
of the potential at
the minimum is not $V(C')$ but, rather, given by Eq.~\eqref{V(C)} as
\be
V(C) = -0.295417 \times 10^2 v^4\, .
\ee

It is true, that given the $\Lambda_4 \rightarrow - \Lambda_4$
interchange between $C$ and $C'$,
the roles of $C$ and $C'$ will be reversed in the case
\ba
&&
M_0 = 13.887500 v^2\, ,
\ \ \ 
\Lambda_0 = 2.928220\, ,
\ \ \ 
\Lambda_1 = -0.305014\, ,
\ \ \ 
\Lambda_2 = -1.018350\, ,
\nonumber\\
&&
\Lambda_3 = 1.365140\, ,
\ \ \ 
\Lambda_4 = +1.053270\, .
\ea
But that is not the issue.
The problem with ignoring $C$ (or $C'$, as the case may be) is that,
for a given set of parameters of the potential, the two candidate
vacua give different values for the potential; one of the
vacua is the global minimum; the other is not.
This issue is even more troublesome if one is looking
for the minima in a completely softly-broken $A_4$ potential,
where such a parameter symmetry is absent,
and one is trying to use the results of the exact 
$A_4$-symmetric potential as a guide.

\subsection{Comparison with the literature}

The expressions for the vevs and masses in cases $A$, $B$, $D$, and $C$
are contained in Ref.~\cite{Degee:2012sk}.
We agree with their results, except that the $\Lambda_4$ term in
their Eq.~(B.10) - their Eq.~(43) on the arxiv version - for the charged Higgs masses in case $C$
must be changed from $\sqrt{3} \Lambda_4/2$ into $\Lambda_4/(2 \sqrt{3})$ in order
to agree with our Eq.~\eqref{mc:C}.

The expressions for the vevs and masses in cases $A$, $B$, $D$, and $C'$
are repeated in Ref.~\cite{deMedeirosVarzielas:2022kbj}, but with the notational change
that $C'$ is renamed $C$ and the appropriate change is made in the sign in front
of $\Lambda_4$. We have found a number of misprints in their
Eqs.~(12), (15), and (16).

The conditions that guarantee that a given vacuum is indeed the absolute minimum
are discussed in the next section.
It turns out that the conditions identified in \cite{Degee:2012sk,deMedeirosVarzielas:2022kbj}
to guarantee that $B$ is the absolute minimum are necessary but not sufficient to that
effect.
In addition,
equations to guarantee that $C'$ and $D$ are the global minimum
are not presented in \cite{Degee:2012sk}.
They are presented in \cite{deMedeirosVarzielas:2022kbj},
where they are wrong.
This is what we turn to next.

\section{\label{sec:BFB-n}Bounded from below conditions (neutral directions)}

The full conditions that guarantee that the exact $A_4$ symmetric
3HDM potential is bounded from below (BFB) are not known.
The problem is that the form of the orbit space for the charged
solutions (solutions of the vev that break electric charge) is not known.
And,
as illustrated in \cite{Faro:2019vcd,Ivanov:2020jra},
guaranteeing BFB only along the neutral directions does \textit{not}
guarantee that the potential is indeed BFB.

The conditions for BFB of the $A_4$ 3HDM along the neutral directions (BFB-n) have been
conjectured in \cite{Ivanov:2020jra},
and proved to hold in
\cite{Buskin:2021eig}.
They are:
\ba
&&
\Lambda_0 + \Lambda_3 \geq 0\, ,
\label{BFB_1}
\\
&&
\frac{4}{3}(\Lambda_0 + \Lambda_3) + \frac{1}{2} (\Lambda_1 + \Lambda_2)
- \Lambda_3
- \frac{1}{2} \sqrt{(\Lambda_1 - \Lambda_2)^2+\Lambda_4^2} \geq 0\, ,
\label{BFB_2}
\\
&&
\Lambda_0 + \frac{1}{2} (\Lambda_1 + \Lambda_2)
+ \frac{1}{2} (\Lambda_1 - \Lambda_2)
\cos{(2 k \pi/3 )}
+ \frac{1}{2} \Lambda_4
\sin{(2 k \pi/3 )} \geq 0\, \ \ (k=1,2,3)\, .
\label{BFB_3}
\ea
But, the left hand side of Eq.~\eqref{BFB_2} is equal to
$4(\Lambda_0+\Lambda_D)/3$. Thus,
the second BFB condition may be traded for
\be
\Lambda_0+\Lambda_D \geq 0\, ,
\label{BFB_xD}
\ee
which holds if and only if Eq.~\eqref{vev:D} representing
$v^2$ at point $D$ is positive.
Similarly,
Eq.~\eqref{BFB_3} with $k=3$ may be written as
\be
\Lambda_0+\Lambda_1 \geq 0\, ,
\label{BFB_xB}
\ee
which according to Eq.~\eqref{vev:B} holds if and only if
the expression for $v^2$ at point $B$ is positive.
Finally, Eq.~\eqref{BFB_3} with $k=1$ and $k=2$ may be written,
respectively, as
\ba
&&
\Lambda_0+ \Lambda_C \geq 0\, ,
\nonumber\\
&&
\Lambda_0 + \Lambda_{C'} \geq 0\, ,
\label{BFB_xCp}
\ea
which according to Eqs.~\eqref{vev:C} and \eqref{vev:C'} hold if and only if
the expressions for $v^2$ at points $C$ and $C'$ are positive,
respectively.
Said otherwise,
the BFB-n conditions of Ref.~\cite{Buskin:2021eig}
may be written compactly as
\be
\Lambda_0 + \Lambda_X \geq 0
\ \ \ \ (\textrm{for all }\, X=A,B,C,C',D)\, .
\label{BFB_better}
\ee
These equations guarantee that the right hand side
of equations \eqref{vev:A}, \eqref{vev:B},
\eqref{vev:C}, \eqref{vev:C'}, \eqref{vev:D}
representing $v^2$ at points $X$ are \textit{all} positive.

To be specific,
when solving the stationarity equations,
one finds the trivial $(0,0,0)$ solution.
One also finds possible solutions at points $P=A,B,C,C',D$,
whenever the right hand side of
\be
v^2 = \frac{\sqrt{3}M_0}{\Lambda_0 + \Lambda_P}
\ee
is positive.
Of course, the $\Lambda$ parameters might be such that
$\Lambda_0 + \Lambda_C >0$,
while $\Lambda_0 + \Lambda_{C'}<0$.
In that case,
$C$ is a critical point but $C'$ is not.
When we mention in the text that we require $v^2 >0$ at a given point $P$,
what we mean is that we require $\Lambda_0 + \Lambda_P >0$.
Eq.~\eqref{BFB_better} can be taken to mean that the potential
is BFB-n if and only if \textit{all} points $P=A,B,C,C',D$
are critical points \textit{simultaneously}. The physical explanation
is discussed in detail in \cite{Ivanov:2020jra};
it hinges on the fact that, for the exact $A_4$ potential,
there is a connection between the BFB
conditions and the value of the quartic part of the potential
at the candidates for global minima.

\section{\label{sec:global-n}Conditions for global minimum}

Recall that requiring that the 125GeV scalar has a positive mass squared
implies that $M_0>0$ for all cases.
Let us assume that we found that a given point $P$ is a local minimum.
Then, in a notation that should be obvious from
Eqs.~\eqref{V(A)},
\eqref{V(B)},
\eqref{V(D)},
\eqref{V(C)},
and \eqref{V(C')},
we conclude that, at that point $P$
one must have $\Lambda_0 + \Lambda_P > 0$.
Let us continue to imagine that $P$ is a local minimum. We wish to know
whether $V(X)$ could be lower at some other point $X \neq P$.
But
\be
V(X) - V(P)
= \frac{M_0^2}{4(\Lambda_0+\Lambda_P)(\Lambda_0 + \Lambda_X)}
\left[\Lambda_X - \Lambda_P
\right]\, .
\label{XP}
\ee
The prefactor on the right hand side of this equation
is guaranteed to be positive by the BFB conditions
in Eq.~\eqref{BFB_better}.
Thus,
$V(X) < V(P)$ can only be attained if it is possible to have
\be
\Lambda_X - \Lambda_P < 0\, ,
\label{V(X)-V(P)_BAD}
\ee
together with $\Lambda_0 + \Lambda_P > 0$,
the other conditions needed for $P$ to be a local minimum,
and the conditions for BFB-n
potential.\footnote{We note in passing that $\Lambda_0 + \Lambda_X > 0$
would be one of the conditions needed in order for $X$ to be a local minimum.}

\subsubsection*{$\mathbf{A:\ (1, 0, 0)}$}

Using the geometrical minimization technique,
Ref.~\cite{Degee:2012sk} identifies the conditions for $(1, 0, 0)$
to be ``stable''
as\footnote{\label{foot:1}The expression ``stable'' can be taken to mean ``local minimum''.
In that case,
Ref.~\cite{deMedeirosVarzielas:2022kbj}, which uses the expression ``global minimum'',
provided a correct generalization. See footnote \ref{foot:2} below.
}
\begin{align}
    \Lambda_3 < 0\, ,
    & &  \Lambda_0 > |\Lambda_3| > -\Lambda_2, -\Lambda_1\, ,
    & & \Lambda_4^2 < 4(\Lambda_1 + |\Lambda_3|)(\Lambda_2 + |\Lambda_3|)\, .
\label{global:A}
\end{align}
It is interesting that \eqref{global:A} can be re-obtained in the following way.
Requiring that $m_{H_i^\pm}^2$ in \eqref{mc:A} is positive,
we obtain $\Lambda_3=-|\Lambda_3|$ is negative.
Thus, from \eqref{vev:A}, $\Lambda_0 > |\Lambda_3|$.
Adding \eqref{mn1:A} and \eqref{mn2:A},
we conclude that $|\Lambda_3|>-(\Lambda_1 + \Lambda_2)/2$.
Requiring that $m_{h_i}^2$ in \eqref{mn1:A} is positive implies that
$(\Lambda_1 + \Lambda_2 + 2 |\Lambda_3|)^2 > (\Lambda_1 - \Lambda_2)^2+\Lambda_4^2$,
from which $\Lambda_4^2 < 4(\Lambda_1 + |\Lambda_3|)(\Lambda_2 + |\Lambda_3|)$.
We wish for $\Lambda^2_4 \neq 0$; $\Lambda_4=0$ would lead to the
$S_4$ symmetric 3HDM.
This coincides almost exactly with \eqref{global:A}.
Defining
\be
y_1 = \Lambda_1 + |\Lambda_3|\, ,
\ \ \ 
y_2 = \Lambda_2 + |\Lambda_3|\, ,
\label{yk}
\ee
the previous conditions require $y_1 + y_2 \equiv \Sigma > 0$ and $y_1 y_2 > 0$.
But this forces both $y_1$ and $y_2$ be larger than $0$ (and, although irrelevant to
our argument, smaller than $\Sigma$, related through \eqref{mn1:A}-\eqref{mn2:A}
with $m_{h_i}^2 + m_{H_i}^2$).
Thus, $|\Lambda_3| > -\Lambda_2, -\Lambda_1$,
and we have recovered all the conditions in
\eqref{global:A}.
A curious fact is that in this argument we did not have to require that
$V(A)$ lie lower than the values of the
potential at the other solutions of the stationarity equation;
$V(B)$, $V(D)$, $V(C)$, $V(C')$.
Indeed, one can show that requiring that, at the vacuum $A$,
$v^2>0$ and that all squared masses are positive\footnote{Strictly speaking,
by this statements we mean that one must require that
the expressions on the right-hand side of Eqs.~\eqref{vev:A}-\eqref{mn2:A}
be positive.}
\textit{guarantees}
that the potential is BFB-n and 
that $V(A) < V(\textrm{other})$.
Said otherwise,
the conditions \eqref{global:A},
which guarantee that $A$ is a local minimum,
also guarantee that it is the global minimum.

We start by proving that \eqref{global:A} imply that the potential is BFB-n.
To this end, we define
\be
x_k = \Lambda_0 + \Lambda_k\ \ \ (k=1,2,3)\, .
\label{xk}
\ee
Eqs.~\eqref{global:A} may be re-expressed as
\begin{align}
    \Lambda_3 < 0\, ,
    & &  x_{1,2,3}>0\, ,
	& &  x_3 < x_1,\, x_2\, ,
    & & \Lambda_4^2 < 4(x_1 - x_3)(x_2 - x_3)\, .
\label{global:ANew}
\end{align}
The second equation \eqref{global:ANew} implies that \eqref{BFB_1} and \eqref{BFB_xB} hold.
We now turn to Eqs.~\eqref{BFB_xCp}. They are satisfied
if
\be
4(\Lambda_0 + \Lambda_C) = x_1 + 3 x_2 \pm \sqrt{3} \Lambda_4 > 0
\ee
is positive.
But,
using the last Eq.~\eqref{global:ANew},
\ba
x_1 + 3 x_2 - \sqrt{3} |\Lambda_4|
&>&
x_1 + 3 x_2 - 2 \sqrt{3} \sqrt{(x_1-x_3)(x_2-x_3)}
\nonumber\\
&=&
(x_1-x_3)+ 3(x_1-x_3) - 2  \sqrt{3(x_1-x_3)(x_2-x_3)} + 4 x_3
\nonumber\\
&=&
\left(
\sqrt{x_1-x_3} - \sqrt{3(x_2-x_3)}
\right)^2
+ 4 x_3 > 0\, ,
\ea
implying that Eq.~\eqref{BFB_xCp} holds.
We now turn to Eqs.~\eqref{BFB_xD}.
We use the last Eq.~\eqref{global:ANew} to show that
\ba
(x_1-x_2)^2 + \Lambda_4^2
&=& \left[ (x_1-x_3)-(x_2-x_3)\right]^2 + \Lambda_4^2
\nonumber\\
&<& \left[ (x_1-x_3)-(x_2-x_3)\right]^2 + 4(x_1-x_2)(x_1-x_3)
\nonumber\\
&=& \left[ (x_1-x_3)+(x_2-x_3)\right]^2\, .
\ea
As a result
\be
\left( x_1 + x_2 + \frac{2}{3} x_3\right)^2 - \left[ (x_1-x_2)^2 + \Lambda_4^2 \right]
>
\left( x_1 + x_2 + \frac{2}{3} x_3\right)^2 - \left(x_1+x_2-2x_3\right)^2 >0\, ,
\ee
leading to
\be
\frac{8}{3}(\Lambda_0 + \Lambda_D)
= x_1 + x_2 + \frac{2}{3} x_3 - \sqrt{(x_1-x_2)^2 + \Lambda_4^2} >0\, ,
\ee
and Eq.~\eqref{BFB_xD} is verified.
We have thus proved that the conditions \eqref{global:A} imply automatically
that the potential is BFB-n.

We now wish to prove that conditions \eqref{global:A} imply that
$\Lambda_{B,C,C',D} > \Lambda_A$,
which,
according to \eqref{XP} means that $A$ is the global minimum.
In fact $\Lambda_B -\Lambda_A = \Lambda_1 + |\Lambda_3| > 0$
is precisely one of Eqs.~\eqref{global:A}.
Using Eqs.~\eqref{Lambda_C}, \eqref{Lambda'_C}, and \eqref{yk},
we find
\ba
4(\Lambda_{C,C'} - \Lambda_A)
&=&
\Lambda_1 + 3 \Lambda_2 -4\Lambda_3 \pm \sqrt{3}\Lambda_4
\nonumber\\
&=&
y_1 + 3 y_2 \pm \sqrt{3} \Lambda_4\, .
\ea
We now use the last Eq.~\eqref{global:A}
to find
\be
(y_1 + 3 y_2)^2 - 3 |\Lambda_4|^2
> (y_1 + 3 y_2)^2 - 12 y_1 y_2 = (y_1 - 3 y_2)^2 >0\, .
\ee
Thus, conditions \eqref{global:A} imply $\Lambda_{C,C'} > \Lambda_A$.
Similarly
\be
\frac{8}{3}(\Lambda_D - \Lambda_A) = y_1 + y_2 - \sqrt{(y_1-y_2)^2 + \Lambda_4^2}\, ,
\ee
and
\be
(y_1+y_2)^2 - (y_1-y_2)^2 -\Lambda_4^2 = 4y_1 y_2 - \Lambda_4^2 >0\, ,
\ee
where the last inequality follows from the last Eq.~\eqref{global:A}.
This completes the proof that asking that $A$ be a local minimum,
as in Eqs.~\eqref{global:A}, automatically guarantees
that the potential is BFB-n and that $A$ is also the global minimum.

We will prove next that this very simple and favorable situation
does not occur for the remaining points of interest.

\subsubsection*{$\mathbf{B:\ (1, 1, 1)}$}

Using the geometrical minimization technique,
Ref.~\cite{Degee:2012sk} identifies the conditions for $(1, 1, 1)$
to be ``stable''
as\footnote{\label{foot:2}The expression ``stable'' can be taken to mean ``local minimum''.
In that case,
the statement in Ref.~\cite{Degee:2012sk} is correct and the statement in
Ref.~\cite{deMedeirosVarzielas:2022kbj}, which uses the expression ``global minimum'',
is incorrect. 
In contrast, if ``stable'' is taken to mean ``global minimum'',
then both Ref.~\cite{Degee:2012sk} and Ref.~\cite{deMedeirosVarzielas:2022kbj}
are incorrect.
In any case, it is best to avoid ``stable'' and use instead the 
unequivocal ``local minimum'' and ``global minimum''.
See also footnote \ref{foot:1} above.
}
\begin{align}
    \Lambda_1 < 0\, ,
    & &  \Lambda_0 > |\Lambda_1| > -\Lambda_2, -\Lambda_3\, ,
    & & \Lambda_4^2 < 12\Lambda_1^2\, ,
    & & \Lambda_4^2 < 2(\Lambda_3 + |\Lambda_1|)(\Lambda_2 + |\Lambda_1|)\, .
\label{global:B}
\end{align}
As in case $A$, we can recover
these conditions by requiring that, at point $B$,
$v^2 >0$ and that all expressions for the squared masses are positive.
Thus, these conditions ensure that $B$ is a local minimum.
These conditions are also enough to prove that
$V(B)< V(A)$.

However,
contrary to what we saw in case $A$,
here the conditions \eqref{global:B} are \textit{not} enough
to guarantee that $V(B) < V(C)$.
Indeed, we have shown that if a point satisfies \eqref{global:B}
but, in addition, satisfies also
\be
\Lambda_0 + \Lambda_C > 0\, ,
\ \ \ \textrm{and}\ \ \ 
\Lambda_4 < 0\, ,
\ \ \ \textrm{and}\ \ \ 
|\Lambda_4| > \sqrt{3}(\Lambda_2 + |\Lambda_1|)\, ,
\ee
then it has $V(C) < V(B)$.
A numeric example is
\ba
&&
\Lambda_0 = 8.8427\, ,
\ \ \ 
\Lambda_1 = -7.36842\, ,
\ \ \ 
\Lambda_2 = -3.62124\, ,
\nonumber\\
&&
\Lambda_3 = 2.62323\, ,
\ \ \ 
\Lambda_4 = -8.03341 < 0\, ,
\ea
which leads to $\Lambda_C=  -8.0366$
(and, thus, $\Lambda_0 + \Lambda_C >0$), and
$|\Lambda_4|-\sqrt{3}(\Lambda_2+|\Lambda_1|) =  1.5431 > 0$.
As a result,
in this case,
\be
\frac{4}{M_0^2} V(B)= -0.678297\, ,
\ \ \ 
\frac{4}{M_0^2} V(C) =  -1.24054 < V(B)\, ,
\ee
even tough Eqs.~\eqref{global:B} are satisfied.

Similarly,
points where Eqs.~\eqref{global:B} are satisfied
but also satisfy
\be
\Lambda_0 + \Lambda_{C'} > 0\, ,
\ \ \ \textrm{and}\ \ \ 
\Lambda_4 > 0\, ,
\ \ \ \textrm{and}\ \ \ 
|\Lambda_4| > \sqrt{3}(\Lambda_2 + |\Lambda_1|)\, ,
\ee
lead to $V(C') < V(B)$.

Finally, we can find points for which
Eqs.~\eqref{global:B} are satisfied but still lead to
$V(B) > V(D)$.
Those are points for which, in addition to \eqref{global:B},
\be
4 \Lambda_0 > 3 \Tilde{\Lambda} - \Lambda_3 > 4 |\Lambda_1|\, .
\ee

\subsubsection*{$\mathbf{C:\ (1, \eta, \eta^{-1})}$}

Ref.~\cite{deMedeirosVarzielas:2022kbj} claims that,
in the region of parameter space where
\begin{align}
    \Lambda_2 < 0 & &  |\Lambda_2| > |\Lambda_1| & & \Lambda_3> \Lambda_1
& & 4\Lambda_0 + 3 \Lambda_1 > 3|\Lambda_2|,
\ \ \textrm{(not correct)}
\label{global:C}
\end{align}
$(1, \eta, \eta^{-1})$ is the global minimum.\footnote{In fact,
Ref.~\cite{deMedeirosVarzielas:2022kbj} is dealing with what we call $C'$;
but their analysis makes no distinction between $C$ and $C'$.}
This is incorrect. We find that these conditions are also not the
correct necessary and sufficient conditions for $C$ to be a local minimum.

To find the correct equations for $C$ to be a local minimum, we proceed as follows.
The first Eq.~\eqref{mc:C} implies that
\be
- \sqrt{3} \Lambda_4  = c> 6 \Lambda_2\, .
\ee
The second Eq.~\eqref{mc:C} implies that
\be
- \sqrt{3} \Lambda_4  = c> \frac{3}{2} (\Lambda_1 + \Lambda_2)\, .
\ee
The last two equations imply that
\be
\Lambda_C < 0\, ,
\ee
a result which can also be obtained by
adding the first and second equations \eqref{mc:C}.
For Eq.~\eqref{vev:C} to yield a positive vev squared,
we must have
\be
\Lambda_0 > - \Lambda_C = |\Lambda_C|\, . 
\ee
Adding the two expressions in Eq.~\eqref{mn:C},
we find
\be
2c > - (a+b) = - (\Lambda_1 - 3 \Lambda_2 + 2 \Lambda_3)
\equiv 2 c_{ab}\, .
\ee
For the first equation \eqref{mn:C} to yield a positive mass squared,
we also need that $A^2 > B^2$.
After some calculations,
this requires that
\be
c < c_- \, \ \ \ \textrm{or}\ \ \ c > c_+ \, ,
\ee
where
\be
2 c_\pm =
-3(a+b) \pm \sqrt{9(a+b)^2-4ab}\, .
\label{c+-}
\ee
However,
one can show that
\be
c_- < c_{ab} < c_+\, .
\ee
Thus,
the solution $c<c_-$ is impossible,
and, moreover,
the simultaneous requirement that $c>c_{ab}$ and $c>c_+$
reduces to
\be
c > c_+\ \ \ \textrm{always}\, .
\ee

The conditions for $C$ to be a local minimum are, thus,
\be
\Lambda_0 > |\Lambda_C|\, ,
\ \ \ 
- \sqrt{3} \Lambda_4 = c > 6 \Lambda_2\, ,
\ \ \ 
c > \frac{3}{2} (\Lambda_1 + \Lambda_2)\, ,
\ \ \ 
c > c_+\, .
\label{local:C}
\ee
We point out that, even after imposing \eqref{local:C},
$\Lambda_1$, $\Lambda_2$, and $\Lambda_4$ can appear with either sign.

It is true that, under the conditions that $A$ be the global minimum
in \eqref{global:A}, it is guaranteed that $V(A) < V(C)$.
So, if $C$ is to be the global minimum,
one or more of the conditions in \eqref{global:A}
must be violated.
We are now interested in what happens to $V(A)-V(C)$
under the conditions \eqref{local:C},
that, as we have just seen, guarantee
that $V(C)$ is a local minimum.
Requiring $V(A) < V(C)$ is only possible if
$\Lambda_0+\Lambda_3>0$, $\Lambda_3 <0$,
and $\Lambda_C - \Lambda_3 >0$.
But the latter condition forces
$c < -(a+2b)$, which is impossible given that $c>c_+$.
We conclude that,
under conditions \eqref{local:C},
$V(C) < V(A)$ always.

In contrast,
requiring $V(B) < V(C)$ is possible if
$\Lambda_0+\Lambda_1>0$, $\Lambda_1 <0$,
and $\Lambda_C - \Lambda_1 >0$.
The latter condition forces
$c < -a$, 
which is only compatible with
$c>c_+$ when $a<0$ and $a+b >0$.
Such points are difficult to generate in a blind scan.
One point obeying 
\eqref{local:C} but having $V(B) < V(C)$ is
\ba
&&
\Lambda_0 = 2.302450\, ,
\ \ \ 
\Lambda_1 = -1.937900\, ,
\ \ \ 
\Lambda_2 = -0.877449\, ,
\nonumber\\
&&
\Lambda_3 = 1.487450\, ,
\ \ \ 
\Lambda_4 = -1.424600\, ,
\ea
which leads to $V(B)-V(C)=  -0.901551 \times (M_0^2/4)$.

We now compare $V(C)$ with $V(C')$.
Given that $\Lambda_{C'} = \Lambda_C + c/2$,
\be
V(C')-V(C)
=
\frac{M_0^2}{8} \frac{1}{\Lambda_0 + \Lambda_C}
\left[
\frac{c}{\Lambda_0 + \Lambda_{C'}}
\right]\, .
\ee
But,
using the BFB conditions in Eq.~\eqref{BFB_better},
we know that $\Lambda_0 + \Lambda_{C}>0$ and $\Lambda_0 + \Lambda_{C'}>0$.
We conclude that,
under conditions \eqref{local:C},
\ba
c > 0 
& \Rightarrow &
V(C') > V(C)\, ,
\\
c < 0 
& \Rightarrow &
V(C') < V(C)\, .
\ea

We now turn to $V(C)$ versus $V(D)$.
Following \eqref{V(X)-V(P)_BAD},
for $V(D) < V(C)$ we would need
$\Lambda_0 + \Lambda_D > 0 $ and
$\Lambda_D - \Lambda_C < 0$,
which, since $\Lambda_C < 0$, requires $\Lambda_D < 0$ and
$\Lambda_0 > |\Lambda_D| > |\Lambda_C|$.
In addition, one can show that
\be
8(\Lambda_D - \Lambda_C)
= A - \sqrt{a^2 + 3 c^2}\, ,
\ee
which would be negative for $V(D) < V(C)$,
despite $ A = a + b + 2c$ defined in \eqref{A} being positive.
We find numerically that these conditions are possible.
An example is,
\ba
&&
\Lambda_0 = 2.695860\, ,
\ \ \ 
\Lambda_1 = 2.176150\, ,
\ \ \ 
\Lambda_2 = -1.581800\, ,
\nonumber\\
&&
\Lambda_3 = 0.585923\, ,
\ \ \ 
\Lambda_4 = -0.856958\, , 
\label{P=C;X=D}
\ea
which leads to $V(D)-V(C)=  -0.0229917 \times (M_0^2/4)$.

\subsubsection*{$\mathbf{C':\ (1, \omega, \omega^2)}$}

The discussion here parallels that on $C$.
One may merely define $c'=+\sqrt{3} \Lambda_4$,
performing the substitution $c \rightarrow c'$ in the formulae of case $C$.
Recall, however,
that, given a numerical point,
$C$ and $C'$ are not ``equivalent''.
Indeed, for a given point with $\Lambda_4 \neq 0$,
$V(C)$ is always different from $V(C')$.

Point $C'$ here is what is named point $C$ in
Ref.~\cite{deMedeirosVarzielas:2022kbj},
where it is claimed that the conditions in
\eqref{global:C} are necessary and sufficient for
$C'$ to be a global minimum.
We have shown that, for example,
the point
\ba
&&
\Lambda_0 = 1.079130\, ,
\ \ \ 
\Lambda_1 = 2.436530\, ,
\ \ \ 
\Lambda_2 =  0.626603\, ,
\nonumber\\
&&
\Lambda_3 =  0.823911\, ,
\ \ \ 
\Lambda_4 = 2.721530\, ,
\ea
corresponds to a potential which is bounded from below,
where $C'$ is both the local and global minimum and which
(in particular because it has $\Lambda_2>0$) violates \eqref{global:C}.

\subsubsection*{$\mathbf{D:\ (1, e^{i\alpha}, 0)}$}

Ref.~\cite{deMedeirosVarzielas:2022kbj} claims that,
in the region of parameter space where
\begin{align}
    \Lambda_2 < 0 & &  |\Lambda_2| > |\Lambda_3| & & \Lambda_1> \Lambda_3
& & 4\Lambda_0 + 3 \Lambda_3 > 3|\Lambda_2|,
\ \ \textrm{(not correct)}
\label{global:D}
\end{align}
$(1, e^{i\alpha}, 0)$ is the global minimum.
This is incorrect. We find that these conditions are also not the
correct necessary and sufficient conditions for $D$ to be a local minimum.
Indeed,
for example,
the point
\ba
&&
\Lambda_0 =  1.178820\, ,
\ \ \ 
\Lambda_1 = -0.137928\, ,
\ \ \ 
\Lambda_2 =  1.458710\, ,
\nonumber\\
&&
\Lambda_3 = -0.607344\, ,
\ \ \ 
\Lambda_4 = -2.660920\, ,
\ea
corresponds to a potential which is bounded from below,
where $D$ is both the local and global minimum and which
(in particular because it has $\Lambda_2>0$) violates
the conditions \eqref{global:D}.

Conversely,
the point
\ba
&&
\Lambda_0 = 1.676240\, ,
\ \ \ 
\Lambda_1 = 0.145391\, ,
\ \ \ 
\Lambda_2 =  -1.943450\, ,
\nonumber\\
&&
\Lambda_3 = -0.044656\, ,
\ \ \ 
\Lambda_4 = -0.358779\, ,
\ea
passes all conditions in \eqref{global:D},
but it is the point $C$ (not the point $D$)
which is the global minimum of the potential.

The correct necessary and sufficient conditions for the point $D$
to be a local minimum can be cumbersome to
write\footnote{In what concerns the $S_4$ 3HDM (not the $A_4$ 3HDM),
Ref.\cite{Degee:2012sk} presents the correct
version of Eq.~\eqref{global:C} and Eq.~\eqref{global:D}.
We have shown that those are indeed
the necessary \textit{and} sufficient conditions for global minima in $S_4$.}.
To simplify their appearance,
we introduce the notation
\be
\ell_k = \Tilde{\Lambda} + \Lambda_k\ \ \ (k=1,2,3)\, .
\label{ell}
\ee
Recall that from the definition of $\Tilde{\Lambda}$ in Eq.~\eqref{Ltilde},
$\ell_1 > 0$, $\ell_2 > 0$ and, thus,
the second Eq.~\eqref{mn1:D} is automatically satisfied.
The first Eq.~\eqref{mc:D} requires $\Tilde{\Lambda}>0$,
the second Eq.~\eqref{mc:D} requires $\ell_3 < 2 \Tilde{\Lambda}$,
and the first Eq.~\eqref{mn1:D} requires $\ell_3>0$.
Eq.~\eqref{mn2:D} may be rewritten as
\be
\frac{4}{v^2} m_{H_i}^2
=
(\ell_1 + \ell_2 - \ell_3)
\pm
(3 \ell_1 - \ell_2) \sqrt{\frac{\ell_2}{\ell_1 + \ell_2}}\, .
\ee
Requiring that these masses squared be positive imposes $\ell_3<L$, where
\be
L = (\ell_1 + \ell_2) - |3 \ell_1 - \ell_2| \sqrt{\frac{\ell_2}{\ell_1 + \ell_2}}\, ,
\label{ell3:def_L}
\ee
and the right hand side is always positive, as one can easily check.
Finally, one must ensure that $\Lambda_0 + \Lambda_D >0$.
As a result,
The conditions for $D$ to be a local minimum are, thus,
\be
\Tilde{\Lambda}>0\, ,
\ \ \ 
0 < \ell_3 < 2 \Tilde{\Lambda}\, ,
\ \ \ 
\ell_3 < L\, 
\ \ \ 
\Lambda_0 > \Tilde{\Lambda} - \frac{\ell_3}{4}\, .
\label{local:D}
\ee
One practical form to simulate points in this approach is to choose
$\Lambda_1$, $\Lambda_2$, and $\Lambda_4$ such that
the first equation holds.
This determines $\ell_1$ and $\ell_2$,
to substitute in Eq.~\eqref{ell3:def_L}.
Then, choose some $\ell_3$
(which determines $\Lambda_3$) satisfying the
next two equations, and finally taking $\Lambda_0$ to obey the last equation.

\subsection{\label{subsec:summary}Neutral global minima: summary}

We are now ready to generate points guaranteed to be the local minimum
in $A_4$ symmetric potentials, which are BFB (at least along the neutral directions).
The \textbf{Procedure} is as follows:
\begin{enumerate}
\item Chose some point $P$ from the list of candidates provided
by \cite{Degee:2012sk}: $\{A,B,C,C',D\}$.
For a given point,
use the respective conditions for local minimum.
These are presented
in Eq.~\eqref{global:A} for point $A$,
in Eq.~\eqref{global:B} for point $B$,
in Eq.~\eqref{local:C} for point $C$,
in Eq.~\eqref{local:C} with $\Lambda_4 \rightarrow - \Lambda_4$ for point $C'$,
and in Eq.~\eqref{local:D} for point $D$.
\item Impose the BFB-n conditions as in Eq.~\eqref{BFB_better}.
\item Guarantee that the chosen point is indeed the global minimum
by imposing the conditions $\Lambda_X -\Lambda_P > 0$, for all $X \neq P$.
\end{enumerate}

This procedure guarantees that all points correspond to potentials that
are BFB-n. But they do not guarantee that the potential is BFB along
the charge breaking directions (BFB-c).
This is what we discuss in the next section.

\section{\label{sec:BFB-c}Protecting against global minima along charged directions}

\subsection{A lower bound for BFB-c}

Using the gauge freedom, one can write the most general vacua
of any 3HDM as \cite{Faro:2019vcd}
\begin{equation}
\label{eq:22}
\langle \phi_1 \rangle
=\sqrt{r_1}\,
  \begin{pmatrix}
    0\\*[1mm]
    1
  \end{pmatrix}\, ,
\quad
\langle \phi_2 \rangle
=\sqrt{r_2}\,
  \begin{pmatrix}
    \sin(\alpha_2)\\*[1mm]
    \cos(\alpha_2) e^{i \beta_2}
  \end{pmatrix}\, ,
\quad
\langle \phi_3 \rangle
=\sqrt{r_3} e^{i \gamma}\,
  \begin{pmatrix}
    \sin(\alpha_3)\\*[1mm]
    \cos(\alpha_3) e^{i \beta_3}
  \end{pmatrix}\, .
\end{equation}
So far, we have discussed what happens to the potential
(evaluated at the vacuum\footnote{Occasionally,
we will not distinguish the field $\phi_i$ from the vev
$\langle \phi_i \rangle$, not the potential $V$
from the evaluation of that potential at some vacuum.
The relevant quantities will be clear from the context.
})
along the neutral directions $\alpha_2=\alpha_3=0$.

But, as shown in \cite{Faro:2019vcd,Ivanov:2020jra},
a potential can be bounded from below along the
neutral directions (BFB-n),
but not along the charge breaking directions (BFB-c).
Unfortunately,
the BFB-c condtions are not yet known for the $A_4$ 3HDM.
So, in this section we develop sufficient conditions for BFB-c,
following the technique developed in \cite{Boto}.

It is convenient to define
\cite{Faro:2019vcd}
\begin{equation}
  \label{eq:25a}
  z_{ij}= (\phi_i^\dagger\phi_i) (\phi_j^\dagger\phi_j)
  - (\phi_i^\dagger\phi_j) (\phi_j^\dagger\phi_i) \quad \text{(no sum)}\, .
\end{equation}
When evaluated at the vevs, it is always true that \cite{Faro:2019vcd}
\begin{equation}
  \label{eq:6}
  0\le z_{ij} \le r_i r_j .
\end{equation}
It is also useful to follow the notation of 
Ivanov and Faro \cite{Faro:2019vcd},
adapted to the $A_4$ 3HDM, and write the quartic potential as
\begin{equation}
\label{eq:34}
  V_4= V_N + V_{CB} + V_{A_4}\, ,
\end{equation}
where
\begin{align}
  \label{eq:5}
  V_N=&\frac{a}{2}\left[(\phi_1^\dagger\phi_1)^2 + (\phi_2^\dagger\phi_2)^2+
    (\phi_3^\dagger\phi_3)^2\right]\nonumber\\
  &+ b \left[(\phi_1^\dagger\phi_1) (\phi_2^\dagger\phi_2) +
    (\phi_2^\dagger\phi_2) (\phi_3^\dagger\phi_3) +
    (\phi_3^\dagger\phi_3) (\phi_1^\dagger\phi_1)
    \right]\, ,
\end{align}
\begin{equation}
  \label{eq:7}
  V_{CB}=c \left( z_{12}+ z_{13}+ z_{23}\right)\, ,
\end{equation}
and
\begin{equation}
  \label{eq:8}
  V_{A_4}= \frac{d}{2}\left\{ e^{- 2 i \alpha}
      \left[ \left( \phi_1^\dagger \phi_2\right)^2
    +\left( \phi_2^\dagger \phi_3\right)^2
    +  \left( \phi_3^\dagger \phi_1\right)^2 \right]
    + \text{h.c.} \right\}\, ,
\end{equation}
where $\alpha$ is defined by Eqs.~\eqref{alpha}.
The relation with \eqref{V_A4} is
\begin{eqnarray}
a= \frac{2}{3}(\Lambda_0 + \Lambda_3)\, ,
& &
b= \frac{1}{3}(2\Lambda_0 - \Lambda_3) + \frac{1}{2} (\Lambda_1 + \Lambda_2)\, ,
\nonumber\\*[1mm]
c= - \frac{1}{2} (\Lambda_1 + \Lambda_2)\, ,
&&
d= \frac{1}{2} \sqrt{(\Lambda_1 - \Lambda_2)^2 + \Lambda_4^2}\, .
\label{eq:11}
\end{eqnarray}

Since the problem of finding necessary
and sufficient conditions for the $A_4$ potential is not yet
solved for the charged directions,
we follow the technique in \cite{Boto} and seek
a lower potential to bound the original potential.
Using the parameterization in Eq.~\eqref{eq:22}, we get
\begin{align}
\label{eq:22b}
V_{A_4}^{Lower}=
-d \left(r_1 r_2 + r_1 r_3 + r_2 r_3\right)\, .
\end{align}
Now we go back to $V_{CB}$. Because of Eq.~(\ref{eq:6}) we always have
\begin{equation}
  \label{eq:30}
  V_{CB} \ge V_{CB}^{\rm Lower}= \min(0,c)\left( r_1 r_2 +
   r_1 r_3 + r_2 r_3 \right)\, .
\end{equation}
Now, using Eq.~(\ref{eq:30}), we can finally write
\begin{equation}
  \label{eq:31}
  V_4 > V_4^{\rm Lower} = V_N + V_{CB}^{\rm Lower} + V_{A_4}^{\rm Lower}\, .
\end{equation}

Therefore this potential bounds the original potential; if it is
BFB, so it is the original one.
We start by defining
\begin{equation}
\label{eq:32}
A =
\begin{pmatrix}
A_d & A_o & A_o\\
A_o & A_d & A_o\\
A_o & A_o & A_d
\end{pmatrix}\, ,
  \end{equation}
where the diagonal ($A_d$) and off-diagonal ($A_0$)
matrix elements are given, respectively by
\begin{eqnarray}
A_d &=& a\ =\ \frac{2}{3}(\Lambda_0+\Lambda_3)\, ,
\nonumber\\[+2mm]
A_o &=& b + \textrm{min}(0,c)-d
\nonumber\\
&=&
\frac{1}{3}(2\Lambda_0 - \Lambda_3)
+\frac{1}{2}(\Lambda_1+\Lambda_2) +
\min(0,-\frac{1}{2}(\Lambda_1+\Lambda_2)) -
\frac{1}{2} \sqrt{(\Lambda_1-\Lambda_2)^2+\Lambda_4^2}\, .
\label{eq:33}
\end{eqnarray}
To check the BFB for $V_4^{\rm Lower}$, 
we just have to check the copositivity of the matrix $A$ \cite{Klimenko:1984qx,Kannike:2012pe}.
This leads to
\be
A_d \geq 0\, ,
\ \ \ \ 
A_d + A_o \geq 0\, ,
\ \ \ \ 
\sqrt{A_d} (A_d + 3 A_0) + \sqrt{2(A_d + A_o)^3} \geq 0\, .
\label{copositivity1}
\ee
The addition of the last equation implies simply that $A_o \geq - A_d/2$.
Thus, the copositivity conditions are
\be
A_d \geq 0\, ,
\ \ \ \ 
A_o \geq -A_d/2\, .
\label{copositivity2}
\ee
Such conditions will ensure sufficient conditions for the potential to be BFB, but they are not necessary.
There will be good points in parameter space that are discarded by this procedure.
We will come to this issue below when we compare the respective sets of points.

\subsection{Numerical simulations and conjectures}

We have made a very large number of simulations of parameters
obeying the BFB-n conditions and the conditions that point $P$ (where $P$ is either
of $A,B,C,C',D$) is a local minimum.
As mentioned above, any point satisfying these conditions with $P=A$ is guaranteed
to have the global minimum at $A$.
In contrast,
points obeying BFB-n and local minimum at $P \in \{B,C,C',D\}$
can have the global minimum at any other point $X$, except $X=A$.
To be specific,
let us generate points obeying BFB-n in Eq.~\eqref{BFB_better}
and the conditions for local minimum at $P=C$ in Eq.~\eqref{local:C}.
Then, we will find points where $C$ is indeed the global minimum.
But, we will also find points where $X \in \{B,C',D\}$ is instead the global minimum.
An example with $P=C$ a local minimum and
a global minimum at $X=D$ below is given in Eq.~\eqref{P=C;X=D}.

Moreover,
BFB-n conditions and the conditions that point $P$ (where $P$ is either
of $A,B,C,C',D$) is a local minimum do not even preclude the existence of
a charge breaking (CB) global minimum below. 
We found such examples when choosing $P=D$ as a local minimum.
One such example is
\ba
&&
\Lambda_0 =  2.92212\, ,
\ \ \ 
\Lambda_1 = -0.276035\, ,
\ \ \
\Lambda_2 =  2.58004\, , 
\nonumber\\
&&
\Lambda_3 = 0.477722\, ,
\ \ \ 
\Lambda_4 = 1.80921\, ,
\label{P=D;X=CB}
\ea
leading to 
\ba
&&
V_A= -0.294131\, ,
\ \ \ 
V_B=  -0.377917\, ,
\ \ \ 
V_C= -0.179483\, ,
\nonumber\\
&&
V_C'= -0.249705\, ,
\ \ \ 
V_D=  -0.379115\, ,
\label{lower_CB}
\ea
in units of $M_0^2/4$.
Thus,
looking only at the neutral minima, 
$D$ would indeed be the global minimum.
However, in this case,
there is a lower lying minimum at a CB point, yielding
$V_\textrm{CB}= -0.38331\,(M_0^2/4)$.
In fact,
the point in Eq.~\eqref{P=D;X=CB} survives even after imposing
the global minimum conditions $\Lambda_X - \Lambda_D >0$
for all neutral $X \in \{ A,B,C,C'\}$.

Starting from \eqref{P=D;X=CB}, one can generate an even worse potential
by gradually lowering $\Lambda_0$ while keeping $\Lambda_{1,2,3,4}$
fixed.\footnote{We are grateful to Igor Ivanov for this remark.}
For a $\Lambda_0$ (roughly) between 0.2844 and 0.313,
all points pass the
conditions for BFB-n and global-$D$, but, nevertheless,
correspond to potentials which are unbounded from below along the CB directions.
For example,
the potential corresponding to
\ba
&&
\Lambda_0 =  0.300000\, ,
\ \ \ 
\Lambda_1 = -0.276035\, ,
\ \ \
\Lambda_2 =  2.58004\, , 
\nonumber\\
&&
\Lambda_3 = 0.477722\, ,
\ \ \ 
\Lambda_4 = 1.80921\, ,
\label{P=D;un-BFB-c}
\ea
obeys BFB-n and the requirement that $D$ be the lowest lying
neutral minimum but, still, is unbounded from below.
This example disproves a conjecture in Ref.~\cite{Ivanov:2020jra}.\footnote{In hindsight,
Ref.~\cite{Ivanov:2020jra} contains in fact two conjectures.
The conjecture that Eqs.~\eqref{BFB_better} plus local-$P$ imply BFB-n,
which was finally proven in \cite{Buskin:2021eig},
and the conjecture that Eqs.~\eqref{BFB_better} plus local-$P$ actually imply
the (stronger) combined BFB-n and BFB-c.
Example \eqref{P=D;un-BFB-c} disproves the latter.}.

But we have found something quite interesting concerning BFB-c in our extensive simulation.
We performed a simulation of over 10000 parameter points, followed by a numerical
minimization searching for the global minimum.
For each of these parameter points, the minimization was repeatedly
attempted with 200 random seed points.
Requiring only that some point $P$ be a local minimum without
requiring BFB-n did yield many cases where the potential was unbounded from
below along the charge breaking directions.
As \eqref{P=D;X=CB} and \eqref{P=D;un-BFB-c} illustrate,
we have also found some points which passed the stronger requirements of
BFB-n with $D$ the lowest lying neutral minimum,
but which yielded potentials either with a lower-lying CB minimum or which were even
unbounded from below along the CB directions.

However, 
in all cases we generated where
BFB-n holds, $P=A,B,C$, or $C'$ is a local minimum,
and where the potential lies lower than at any other neutral point
$X \neq P$, we found no lower lying CB minimum
and no potential unbounded from below along CB directions.
We propose this result here in the form of a \textbf{Conjecture}:
\begin{itemize}
\item[] The combined requirements that BFB-n holds and that point $P$ (where $P$ is either
of $A,B,C,C'$) is a global minimum \textit{guarantee} that the potential is also BFB-c
and also that there in \textit{no} lower-lying CB global minimum.
\end{itemize}
The analytical proof of this conjecture remains as an open difficult problem.

\section{\label{sec:concl}Conclusions}

We have revisited the conditions for global minimum in the exact $A_4$
3HDM, in light of the candidates for local minimum identified
in \cite{Degee:2012sk} via the geometric minimization method.
We stressed that $C$ and $C'$ are not equivalent,
since $V(C) \neq V(C')$.
Given a set of $\Lambda$ parameters,
both must be probed as candidates for global minimum.
For point $A$,
we found that the conditions determined in \cite{Degee:2012sk} and repeated
in \cite{deMedeirosVarzielas:2022kbj} are indeed conditions that
guarantee that $A$ is the global minimum.
In fact, we have found that, by themselves,
they also imply that the potential is BFB-n.
The conditions for local/global minimum at points $B$, $C$, $C'$, and $D$
are either absent or wrong in the literature
\cite{Degee:2012sk,deMedeirosVarzielas:2022kbj}.
We present here the correct conditions
for $B$, $C$, $C'$, and $D$ to be the global minimum.

The clarification provided here as to the number of global minima
candidates and the exact conditions for each global minimum
is also important given recent developments
proposing specific types of soft-symmetry breaking which preserve de form
($A,B,C,C'$ or $D$) of the critical points which are candidates for
global minimum
\cite{deMedeirosVarzielas:2021zqs,deMedeirosVarzielas:2022kbj}.
In a numerical simulation,
the correct conditions with an exact symmetry can be used to gain confidence
in the results with soft-symmetry breaking,
as the $A_4$ soft-breaking terms are made smaller.

Finally, we address the issue of the charge breaking directions.
We present a sufficient analytical condition for the potential
to be BFB-c. We also perform an extensive numerical simulation,
resulting in a conjecture about BFB-c,
proposing that,
after BFB-n and global-$P$,
only points with $P=D$ are at risk of allowing for
a lower-lying CB (true) global minimum or unbounded from below potentials.

\vspace{5ex}

{\Large \textbf{Acknowledgments}}

\noindent
We are very grateful to Igor Ivanov for helpful discussions,
suggestions,
and a careful reading of the manuscript.
We are also grateful to Ivo de Medeiros Varzielas for
useful comments. 
This work is supported in part by FCT under Contracts
CERN/FIS-PAR/0008/2019,
PTDC/FIS-PAR/29436/2017,
UIDB/00777/2020,
and UIDP/00777/2020;
these projects are partially funded through POCTI (FEDER),
COMPETE,
QREN,
and the EU.



\end{document}